\documentclass[preprint,12pt]{elsarticle}

\usepackage{amssymb}
\usepackage{amsmath}
\usepackage{orcidlink}
\usepackage{xurl}
\usepackage{pdflscape} 
\usepackage{multirow}

\usepackage{xcolor} 
\definecolor{LightGray}{gray}{0.9}

\journal{Original Software Publications (OSP)}

\begin{document}

\begin{frontmatter}



\title{PANTHER: Pluginizable Testing Environment for Network Protocols} 


\author[a]{Christophe Crochet \orcidlink{0000-0001-8635-2098}}

\author[a]{John Aoga \orcidlink{0000-0002-7213-146X}}


\affiliation[a]{organization={UCLouvain},
            email={\{christophe.crochet, john.aoga\}@uclouvain.be},
            country={Belgium}}
            
\author[b]{Axel Legay \orcidlink{0000-0003-2287-8925}}
\affiliation[b]{organization={Legay Consulting},
            email={axellegay@gmail.com},
            country={Belgium}}

\begin{abstract}
In this paper, we introduce \texttt{PANTHER}, a modular framework for testing network protocols and formally verifying their specification. The framework incorporates a plugin architecture to enhance flexibility and extensibility for diverse testing scenarios, facilitate reproducible and scalable experiments leveraging Ivy and Shadow, and improve testing efficiency by enabling automated workflows through YAML-based configuration management. Its modular design validates complex protocol properties, adapts to dynamic behaviors, and facilitates seamless plugin integration for scalability. Moreover, the framework enables a stateful fuzzer plugin to enhance implementation robustness checks.
\end{abstract}

\end{frontmatter}

\begin{table}[h!]
\resizebox{\columnwidth}{!}{%
\centering
\begin{tabular}{|c|l|l|}
\hline
\textbf{Nr.} & \textbf{Code metadata descryption}                                                                       & \textbf{Please fill in this column} \\ \hline
C1           & Current code version                                                                                     & v1.0.3                              \\ \hline
C2           & \begin{tabular}[c]{@{}l@{}}Permanent link to code/reposi-\\ tory used for this code version\end{tabular} & \url{https://github.com/ElNiak/PANTHER}   \\ \hline
C3           & \begin{tabular}[c]{@{}l@{}}Permanent link to Reproducible\\ Capsule\end{tabular}                         &                                     \\ \hline
C4           & Legal Code License                                                                                       & MIT                                 \\ \hline
C5           & Code versioning system used                                                                              & git                                 \\ \hline
C6 &
  \begin{tabular}[c]{@{}l@{}}Software code languages, tools,\\ and services used\end{tabular} &
  \begin{tabular}[c]{@{}l@{}}Python 3.10, docker, Ivy, Shadow\end{tabular} \\ \hline
C7 &
  \begin{tabular}[c]{@{}l@{}}Compilation requirements, oper-\\ ating environments and depen-\\ dencies\end{tabular} &
  \begin{tabular}[c]{@{}l@{}}Linux OS (Tested on Ubuntu, Debian);\\ Docker version 27.2.1,  build 9e34c9b\end{tabular} \\ \hline
C8           & \begin{tabular}[c]{@{}l@{}}If available, link to developer\\ documentation/manual\end{tabular}           & \url{https://elniak.github.io/PANTHER}    \\ \hline
C9           & Support email for questions                                                                              & christophe.crochet@uclouvain.be     \\ \hline
\end{tabular}
}
\caption{Table 1: Code metadata (mandatory)}

\end{table}

\section{Motivation and significance}

Modern network protocols are vital for the reliable operation of distributed systems. However, the increasing complexity and heterogeneity of network protocols present significant challenges for testing and verification. Dynamic behaviors, time-varying properties, and the unpredictability of real-world conditions require robust methodologies that extend beyond traditional approaches like Model-Based Testing (MBT) or interoperability testing. These limitations are particularly evident in evaluating protocols’ timing-sensitive features, such as congestion control and retransmissions, which demand both precision and reproducibility.

Tools like NS2 and NS3~\cite{riley2010ns} offer robust simulation environments but lack support for executing real protocol implementations. This limits their applicability in evaluating real-world scenarios and dynamic behaviors. Frameworks like TorXakis~\cite{tretmans2019model} integrate formal methods for network protocol validation. However, these approaches often fail to accommodate extensibility and reproducibility, critical for testing modern protocols with diverse requirements. While tools like those used for \textit{QUIC} testing focus on specific protocols, they lack a generalized architecture for broader applicability.

Network Simulator-centric Compositional Testing (NSCT) \cite{nsct24} methodology is introduced as a significant advance, integrating formal tools like \texttt{Ivy} \cite{Padon_McMillan_Panda_Sagiv_Shoham_2016,McMillan_Padon_2020} with deterministic network simulators such as \texttt{Shadow} \cite{jansen2011shadow,280766}. NSCT showcased the effectiveness of the methodology test the real word QUIC protocol. However, NSCT's scope was limited, lacking the necessary extensibility for broader adoption and experimentation with diverse protocols. \texttt{PANTHER}~\cite{panther} is a tool implementing of NSCT that overcomes these limitations.

\texttt{PANTHER} is a modular framework that combines formal verification and dynamic simulations to validate network protocol correctness. It incorporates a plugin architecture to enhance flexibility and extensibility for diverse testing scenarios, facilitates reproducible and scalable experiments leveraging Ivy and Shadow, and improves testing efficiency by enabling automated workflows through YAML-based configuration management. By enabling users to configure experiments according to specific needs, \texttt{PANTHER} allows for precise validation of both functional and non-functional properties of protocols.

The remainder of this paper is structured as follows: Section 2 provides a detailed overview of \texttt{PANTHER}’s architecture and methodology. Section 3 presents an illustrative example, and Section 4 the impact of the tool.

\section{Software description}

\subsection{Overview of PANTHER}

\texttt{PANTHER} is constructed with a modular framework aimed at supporting both extensibility and reproducibility. Its main elements consist of testers, such as the adversarial testing modules generated by \texttt{Ivy} for formal verification; implementation components under study (IUTs) like QUIC and MiniP; execution environments that use runtime analysis tools; and network setups employing \textit{Shadow} for deterministic simulations with precise network parameters. \texttt{PANTHER} incorporates \texttt{Ivy}, a formal verification utility that uses a domain-specific language (\texttt{.ivy}) to articulate protocol specifications for formal requirements. \texttt{Ivy} processes these formal models to produce executable \texttt{C++} testers, designed as adversarial test modules that methodically explore the protocol's state space, verifying adherence to the established specifications.

The framework emphasizes reproducibility, using the \texttt{Shadow} network simulator to provide deterministic simulations of real-world conditions. \texttt{Shadow} allows precise control over parameters like latency, jitter, and bandwidth, enabling consistent experiment replication, especially for time-sensitive protocols like retransmissions and congestion control, everything encapsulated and orchestrated in a Docker container build automatically.

Additionally, \texttt{PANTHER} includes a self-contained Python package along with clear documentation, making it accessible and ready for use by the broader community.

\subsection{Software architecture}
\texttt{PANTHER}’s plugin-based architecture enhances modularity and extensibility, supporting plugins for testing modules, Implementation Under Test (IUTs), and environments. This architecture facilitates the seamless incorporation of novel elements, such as protocol-specific IUTs or bespoke settings, permitting \texttt{PANTHER} to adjust to changing testing requirements without significant re-engineering. At the core of the framework there are two main components: plugins and a configuration management system.

\paragraph{Plugins} \texttt{PANTHER} comprises three categories of plugins:  services, protocols and environments. These plugins enable users to specify schemas, testing requirements, and command templates. They are dynamically incorporated via a \texttt{PluginManager}. Table~\ref{tab:plugin_categories} presents these plugins.

\begin{table}[h!]
\centering
\begin{tabular}{|p{2.2cm}|p{2cm}|p{8cm}|}
\hline
\textbf{Category} & \textbf{Type} & \textbf{Description} \\
\hline
\multirow{2}{*}{Services} & Testers & Testers generate, execute, and analyze tests for both formal verification and adversarial scenario. \\
\cline{2-3}
& IUTs & IUT plugins define the implementation of protocols (e.g.,\texttt{picoquic}), its validation logic. A \texttt{Jinja} template is required to automate the generation of commands. \\
\hline
\multirow{2}{*}{Environment} & Execution & Execution environments manage runtime contexts for experiments, integrating tools like \texttt{strace} (tracing) and \texttt{gperf} (profiling). \\
\cline{2-3}
& Network & Network environments simulate conditions using tools such as \texttt{Shadow} for deterministic simulations or Docker Compose for multi-container setups. \\
\hline
Protocol & Communi- cation Models & Protocol plugins define communication models (e.g., client-server) and protocol-specific features (e.g., CIDs for \texttt{QUIC}). \\
\hline
\end{tabular}
\caption{Overview of Plugins supported by \textit{PANTHER}.}
\label{tab:plugin_categories}
\end{table}

\paragraph{Configuration Management and Experiment Setup}
The configuration management system enables users to define experiments in YAML files. Configurations specify the protocol's IUT, the tester (e.g., \texttt{Ivy}), the network environment (e.g., \texttt{Shadow} with latency and bandwidth), and execution environment settings (e.g., \texttt{gperf}). A \texttt{ConfigLoader} processes these configurations, validates them against schemas provided by plugins. The validated configurations are then dynamically processed using \texttt{Jinja2} templates to generate experiment setups, such as Docker Compose files, \texttt{Shadow} configurations, and service launch commands, ensuring flexibility and reproducibility.

\subsection{Software workflow}

\begin{figure}[h!]
  \centering
  \includegraphics[width=0.95\columnwidth]{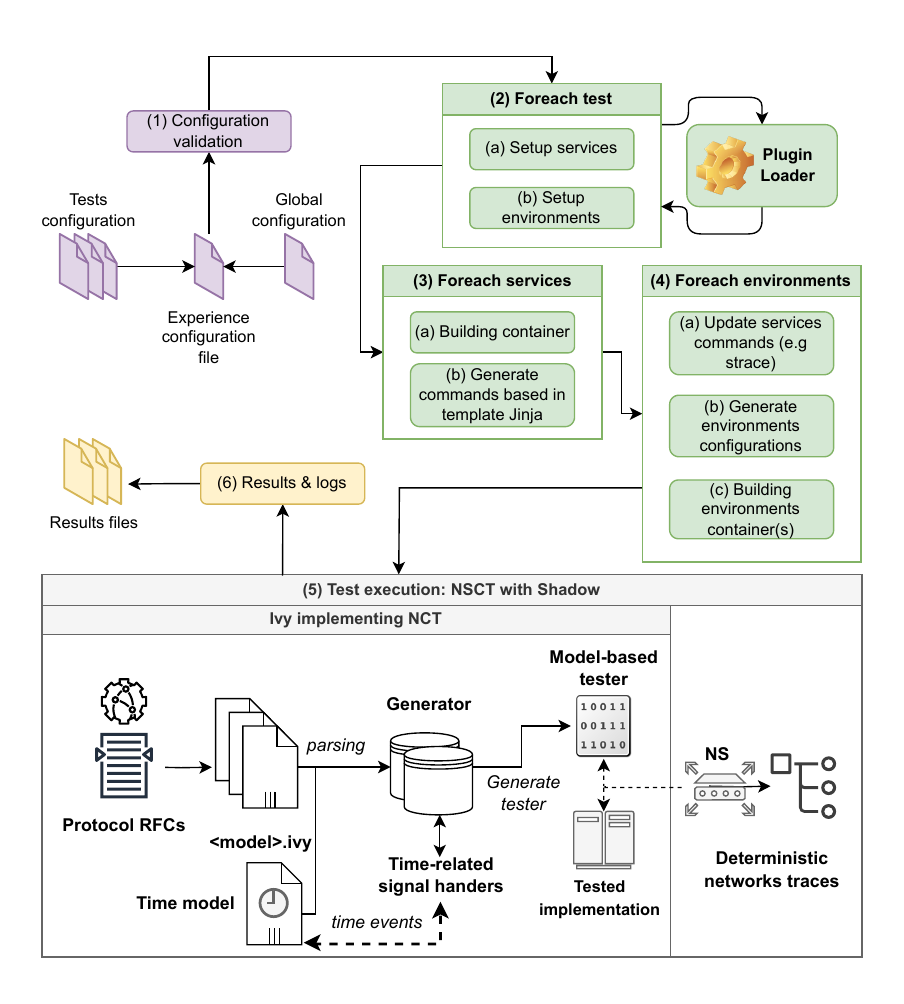}
  \caption{\texttt{PANTHER} workflow}
  \label{fig:workflow}
\end{figure}

The \texttt{PANTHER} Workflow diagram (Figure~\ref{fig:workflow}) outlines the process for testing network protocols. Initially, it involves validating configurations, then deploying services and environments via plugins. To maintain uniform execution, services are containerized, and dynamic service commands are crafted using \texttt{Jinja} templates. Subsequently, environments are set up, and containers built to ensure repeatability. Protocol specifications enable Ivy to generate model-based testers, concentrating on timing aspects like retransmissions and congestion control. Using \texttt{NSCT} with the Shadow simulator guarantees deterministic network simulations and captures critical timing events during testing. Post-testing, deterministic network traces and results are provided, yielding comprehensive logs for evaluation.



\section{Illustrative example}

Our artifact contains many experiment configurations files demonstrating how to launch experiments with a detailed documentations. Additionally, We included many tutorials on how to add new plugins for each categories.

\section{Impact}

\texttt{PANTHER} uniquely combines formal verification with realistic, reproducible simulations, allowing testing of actual protocol implementations. In contrast to NS2 and NS3 \cite{riley2010ns} that lack native execution of real code, \texttt{PANTHER} uses \textit{Shadow} to offer deterministic simulations, precisely managing network parameters like \textit{latency} and \textit{jitter}. Shadow accelerates formal verification and ensures time-dependent safety properties \cite{nsct24}. Furthermore, \texttt{PANTHER} utilizes \texttt{Ivy}’s black-box approach to validate protocols against formal specifications \cite{Crochet_Rousseaux_Piraux_Sambon_Legay_2021}. Its plugin architecture offers extensibility for new protocols, environments, and modules, supporting multiple protocols in a single experiment.

\section{Future work}

Planned upgrades for \texttt{PANTHER} involve a graphical user interface (GUI) for result visualization and experiment design, allowing intuitive visual network scenario configurations to facilitate setup and analysis. A stateful fuzzer plugin is also under development to augment \texttt{Ivy}’s formal verification, aimed at probing protocol state transitions to identify vulnerabilities and enhance implementation robustness checks. Moreover, our formal attack framework will be integrated to harmonize specifications, testing modules, and network environments, thereby improving \texttt{PANTHER}’s proficiency in validating implementations against specifications \cite{crochet2024formally}.

\paragraph{\textbf{Acknoledgements}} We would like to thank the belgium's "\textit{Defence-related Research Action}" (DEFRA) and the "\textit{Automated Methodology for Common Criteria Certification}" project (AMC3) \cite{amc3}.

\bibliographystyle{splncs04}
\bibliography{refs}

\end{document}